\input harvmac
\input epsf

\Title{\vbox{\baselineskip12pt
\hbox{CERN-TH/96-41}
\hbox{UCLA/96/TEP/6}
\hbox{McGill/96-05}
\hbox{hep-th/9602102}
}}
{\vbox{\centerline{$M$-Theory on a Calabi-Yau Manifold}}}

\centerline{Sergio Ferrara$^{a,b}$\footnote{$\dagger$}{Supported in 
part DOE grant DE-FG03-91ER40662, Task C., by EEC Science Program 
SC1$^*$CT92-0789 and by I.N.F.N.}, Ramzi R.~Khuri$^{a,c}$ 
and Ruben Minasian$^a$} \bigskip\centerline{$^a${\it CERN,
 CH-1211, Geneva 23, Switzerland}}
\bigskip\centerline{$^b${\it Physics Department, UCLA, Los Angeles, CA, 
USA}} 
\bigskip\centerline{$^c${\it Physics Department, McGill
University, Montreal, PQ, H3A 2T8 Canada}}
\vskip .3in

We compactify $M$-theory on a Calabi-Yau manifold to five dimensions
by wrapping the membrane and fivebrane solitons of the eleven-dimensional
supergravity limit around Calabi-Yau two-cycles and four-cycles respectively.
We identify the perturbative and non-perturbative BPS states thus obtained
with those of heterotic string theory compactified on $K3\times S^1$. Quantum
aspects of the five-dimensional theory are discussed.

\vskip .3in
\Date{\vbox{\baselineskip12pt
\hbox{CERN-TH/96-41}
\hbox{UCLA/96/TEP/6}
\hbox{McGill/96-05}
\hbox{February 1996}}}

\def\sqr#1#2{{\vbox{\hrule height.#2pt\hbox{\vrule width
.#2pt height#1pt \kern#1pt\vrule width.#2pt}\hrule height.#2pt}}}

\lref\asp{P.~S.~Aspinwall and J.~Louis, hep-th/9510234.}

\lref\vafawit{C.~Vafa, E. Witten, Nucl. Phys. {\bf B447} (1995) 261.}

\lref\stst{M.~J.~Duff and R.~R.~Khuri, Nucl. Phys. {\bf B411}
(1994) 473.}

\lref\dufflu{M.~J.~Duff and J.~X.~Lu, Nucl. Phys. {\bf B354} (1991) 129.}

\lref\duflufb{M.~J.~Duff and J.~X.~Lu, Nucl. Phys. {\bf B354} (1991) 141.}

\lref\duffsw{M.~J.~Duff, Nucl. Phys. {\bf B442} (1995) 47.}

\lref\duffem{M.~J.~Duff, hep-th/9509106.}

\lref\duffm{M.~J.~Duff and R.~Minasian, Nucl. Phys. {\bf B436} (1995) 507.}

\lref\dufflm{M.~J.~Duff, J.~T.~Liu and R.~Minasian, Nucl. Phys. 
{\bf B452} (1995) 261.}

\lref\witfb{E.~Witten, hep-th/9512062.}

\lref\duffmw{M.~J.~Duff, R.~Minasian and E.~Witten, hep-th/9601036.}

\lref\hult{C.~M.~Hull and P.~K.~Townsend, Nucl. Phys. {\bf B438} (1995) 109.}

\lref\wit{E.~Witten, Nucl. Phys. {\bf B443} (1995) 85.}

\lref\schwarz{J.~H.~Schwarz, hep-th/9510086; hep-th/9601077.}

\lref\horw{P.~Ho\v rava and E.~Witten, Nucl. Phys. {\bf B460} (1996) 506.}

\lref\witsi{E.~Witten, Nucl. Phys. {\bf B460} (1996) 541.}

\lref\antft{I.~Antoniadis, S.~Ferrara and T.~R.~Taylor, Nucl. Phys.
{\bf B460} (1996) 489.}

\lref\becker{K.~Becker, M.~Becker and A.~Strominger, Nucl. Phys.
{\bf B456} (1995) 130.}

\lref\cadcdf{A.~C.~Cadavid, A.~Ceresole, R.~D'Auria and S.~Ferrara,
Phys. Lett. {\bf B357} (1995) 76.}

\lref\prep{M.~J.~Duff, R.~R.~Khuri and J.~X.~Lu,
Phys. Rep. {\bf 259} (1995) 213.}

\lref\dkl{M.~J.~Duff, R.~R.~Khuri and J.~X.~Lu,
Nucl. Phys. {\bf B377} (1992) 281.}

\lref\hmono{R.~R.~Khuri, Phys. Lett. {\bf B259} (1991) 261;
Nucl. Phys. {\bf B387} (1992) 315.}

\lref\gst{M.~G\"{u}naydin, G.~Sierra and P.~K.~Townsend, Nucl. Phys. 
{\bf 242} (1984) 244.}

\lref\cardoso{G.~L.~Cardoso, G.~Curio, D.~L\"{u}st, T.~Mohaupt and S.-J.~Rey,
hep-th/9512129.}

\lref\kap{V.~Kaplunovsky, J.~Louis and S.~Theisen, Phys. Lett.
{\bf B357} (1995) 71.}

\lref\hly{S.~Hosono, B.~H.~Lian and S.~T.~Yau, alg-geom/9511001.}

\lref\dufkmr{M.~J.~Duff, R.~R.~Khuri,
R.~Minasian and J. Rahmfeld, Nucl. Phys. {\bf B418} (1994) 195.}

\lref\dabghr{A.~Dabholkar, G.~Gibbons, J.~A.~Harvey and F.~Ruiz
Ruiz, Nucl. Phys. {\bf B340} (1990) 33.}

\lref\town{P.~K.~Townsend, hep-th/9512062.}

\lref\papa{G.~Papadopoulos and P.~K.~Townsend, Phys. Lett. {\bf B357} (1995) 
300.}

\lref\hosono{S.~Hosono, A.~Klemm, S.~Theisen and S.~T.~Yau, 
Comm. Math. Phys. {\bf 167} (1995) 301.}

\lref\kachru{S.~Kachru and C.~Vafa, Nucl. Phys. {\bf B450} (1995) 69.}

\lref\mirror{S.~Ferrara, J.~A.~Harvey, A.~Strominger and C.~Vafa,
Phys. Lett. {\bf B361} (1995) 59.}

\lref\bers{M.~Bershadsky, S.~Cecotti, H.~Ooguri and C.~Vafa,
Nucl. Phys. {\bf B405} (1993) 279.}

\lref\cardtwo{G.~Curio, Phys. Lett. {\bf B368} (1996) 78.}

\lref\antgnt{I.~Antoniadis, E.~Gava, K.~S.~Narain and T.~R.~Taylor, 
hep-th/9507115.}

\lref\klemm{A.~Klemm, W.~Lerche and P.~Mayr, Phys. Lett. 
{\bf B357} (1995) 313.}

\lref\witcy{E.~Witten, hep-th/9602070.}


\newsec{Introduction}

The existence of an underlying eleven-dimensional theory
(the so-called $M$-theory 
\refs{\wit\dufflm\schwarz\horw\witfb{--}\duffmw})
whose low-energy limit is eleven-dimensional supergravity is crucial
to the establishment of the various string/string dualities
recently studied
\refs{\stst\duffm{--}\hult,\wit,\duffsw,\duffmw}.
In this framework, the five seemingly distinct string theories
arise as weak coupling limits of the various compactifications of the
eleven-dimensional $M$ theory, in which the membrane and fivebrane that
naturally arise are either wrapped around or reduced on the compactified
directions.

Although string/string dualities have best been seen in $D=6$, the 
generalization of electric/magnetic duality of super Yang-Mills field
theories requires an $N=2$ duality in $D=4$, which entails a duality (second 
quantized mirror symmetry \mirror) between the heterotic string 
on $K3\times T^2$ and the type $IIA$ string on a Calabi-Yau
threefold \kachru. Several checks of the latter have recently
been carried out \refs{\kap,\antgnt}. It was further observed that this
duality can be lifted to five dimensions to relate the heterotic string on
$K3\times S^1$ and $M$-theory on a Calabi-Yau \antft. This can be seen as
the decompactification limit of the $D=4$ theory when the CY volume becomes
large. When the CY manifold is a $K3$ fibration \refs{\klemm,\asp},
classical calculations in $M$-theory can be matched with one-loop 
calculations on the heterotic side. 
Further evidence in support of this duality can be seen
through the matching of string and point-like fundamental and solitonic
states and through one-loop tests along the lines of
\refs{\duffm,\vafawit,\dufflm}. The fundamental heterotic string state 
arises 
from the $M$-theory fivebrane wrapped around a four-cycle in the CY space,
while the point-like solitonic state resulting from the wrapping of
the heterotic fivebrane around $K3\times S^1$ arises from the $M$-theory
membrane wrapped around a two-cycle in the CY space.
A further reduction to four-dimensions
yields the usual electric/magnetic duality.
Connections are also made with ten-dimesional
type $IIA$ membrane/fourbane duality and to six-dimensional 
heterotic/heterotic
duality.

An outline of this paper is as follows: In section 2 we identify the 
perturbative BPS states in $D=5$ obtained from $M$-theory compactified on a
Calabi-Yau manifold, the string-like and point-like states arising as pairs
of electric/magnetic duals. In section 3 we show that 
for $M$-theory compactified on a specific Calabi-Yau manifold with
$h_{(1,1)}=3$ and $h_{(2,1)}=243$ this electric/magnetic duality follows
from six-dimensional heterotic/heterotic duality. Furthermore, these string
and point-like states can 
also arise from the heterotic string
or the heterotic fivebrane compactified on $K3\times S^1$.
A new vector-gravity interaction is derived in section 4,
providing a one-loop test of the five-dimensional duality. Finally,
in section 5, we discuss quantum aspects of the duality in five dimensions.
In particular, we show that the gauge and gravitational anomalies of the
bulk lagrangian in presence of string-like excitations are cancelled
by the anomalous variation of a boundary term of a chiral worldsheet
string action.

\newsec{$M$-Theory on Calabi-Yau}

It has often been the case that the first manifestation of 
a duality is the exchange of perturbative and non-perturbative states, 
represented by fundamental and solitonic classical solutions (see \prep\ 
and references therein). In establishing the dualities between 
$M$-theory and the various string theories, it is necessary to
investigate the states obtained
after compactification from the solitonic membrane
and fivebrane solutions of the eleven-dimensional supergravity
low-energy limit \hult. 
In compactifying $M$-theory on 
a Calabi-Yau manifold to an $N=2$ supersymmetric theory in
five dimensions \refs{\cadcdf,\papa}, the membrane and fivebrane
wrapped around two- and four-cycles of the Calabi-Yau space
give rise to BPS states in $D=5$.\foot{Note that ``wrapping'' a $p$-brane around a manifold entails simultaneously
compactifying spacetime and its worldvolume 
on that manifold, while ``reducing''  a $p$-brane on a manifold
entails no worldvolume compactification. So a string
wrapped around $S^1$, for example,
 yields a point-like object in the lower dimension,
while a string reduced on $S^1$ remains a string in the lower dimension.}

In \antft, the conjecture was made that the effective theory of
heterotic string theory compactified on $K3\times S^1$ is dual to
eleven-dimensional supergravity compactified on a Calabi-Yau threefold.  
This theory is also equivalent to type $IIA$ string theory compactified
on the same Calabi-Yau threefold, in an appropriate large volume limit.
Quantum effects in five dimensions were also studied \antft.

Following \antft, point-like (electric) states
are obtained in $D=5$ by wrapping the membrane from $M$-theory
around two-cycles in
the Calabi-Yau space. Denote two-cycles and four-cycles respectively
by $C^{2\Lambda}$ and $C_{4\Lambda}$, where $\Lambda=1,...,h_{(1,1)}$.
The charges of these states are obtained from the charge of the membrane by

\eqn\emem{e_\Lambda=\int_{C_{4\Lambda}\times S^3} G_7,}
where $G_7={\delta {\cal L} \over \delta F_4}$, where $F_4=dA_3$ is the 
field strength of the three-form antisymmetric tensor field.

String-like (magnetic) states in $D=5$ arise by wrapping the fivebrane
around four-cycles in the Calabi-Yau space. The charges of these states
are then obtained from the charge of the fivebrane by

\eqn\mfiv{m^\Lambda=\int_{C^{2\Lambda}\times S^2} F_4.}

These states contribute to the point-like and string-like
central charges in $D=5$ via \antft
\eqn\cencha{\eqalign{Z_e&=\sum_\Lambda t^\Lambda e_\Lambda,\cr
Z_m&=\sum_\Lambda t_\Lambda m^\Lambda,\cr}}
where $t^\Lambda$ are the $D=5$ special coordinates and 
$t_\Lambda=C_{\Lambda \Sigma \Delta} t^\Sigma t^\Delta$ are the ``dual'' 
coordinates, $C_{\Lambda \Sigma \Delta}$ being the CY topological 
intersection matrix.

Since the membrane and fivebrane are electric/magnetic duals in eleven
dimensions, the above point-like and string-like states are dual to each 
other in the electric/magnetic sense and correspond to point-like and
string-like soliton solutions \hult. 

A further test of this duality can be performed
in a straightforward manner as follows: a given point-like solution, when 
viewed as a solution of the point-like supergravity
theory in $D=5$, should appear to
be singular and require the addition of a sigma-model source action to
compensate the singularity. From the dual (string) viewpoint, the point-like
solution should appear nonsingular. Similarly, a string solution should appear
singular from the point of view of the string theory in $D=5$ but nonsingular
from the dual, point-like viewpoint. 

Singularity of a solution in a given theory is tested by probing 
the solution with a test-probe which is a fundamental object of the theory
\dkl.
If the probe reaches the origin in finite proper time, the solution is
deemed singular with respect to the theory. If the probe takes an infinite
proper time to reach the source, then the solution is considered nonsingular,
as no singularity can be observed in finite proper time. 
For example, the point-like solution
obtained by wrapping the membrane around a two-cycle should appear 
singular when viewed by a test point-object of the point-like theory in $D=5$, but nonsingular when viewed by a test string of the dual string theory in 
$D=5$.

In fact, the singularity criteria for the electric/magnetic dual objects
at hand can be seen to be satisfied immediately in $D=5$, since all objects 
in question
are point-like or strings, and it was shown in \prep\ that provided at least
one of the two objects in question is either a string or a point, then it
is self-singular and mutually nonsingular with its dual.

\newsec{Five-Dimensional Duality}

In a recent paper \duffmw, heterotic string/string duality was
examined from the point of view of $M$-theory, where it was argued that the $E_8\times E_8$ heterotic string compactified on $K3$ with equal instanton
 numbers in the two $E_8$'s is
self-dual, a result which can be seen by looking in two different ways at
eleven-dimensional $M$-theory compactified on $K3\times S^1/Z_2$. 
One weakly coupled heterotic string is obtained by wrapping the $D=11$ 
membrane around 
$S^1/Z_2$, while the dual heterotic string, also weakly coupled, is obtained by reducing the
$D=11$ fivebrane on $S^1/Z_2$ and then wrapping around $K3$. Each of these
two strings is strongly coupled from the point of view of the dual one.

If we further compactify by reducing the first six-dimensional
heterotic string on $S^1$ and wrapping the dual six-dimensional heterotic
string on $S^1$, we obtain on the one hand a string in five dimensions
and on the other a dual, point-like object in five dimensions.
We claim that, starting with a $K3$ vacuum in which the gauge symmetry is 
completely Higgsed, 
this $D=5$ string can be identified with the $M$-theory
fivebrane wrapped around a Calabi-Yau four-cycle, while the $D=5$ point-like
object can be identified with the $M$-theory membrane wrapped 
around a Calabi-Yau two-cycle for the specific Calabi-Yau manifold
$X_{24}(1,1,2,8,12)$ with $h_{(1,1)}=3$ and $h_{(2,1)}=243$ 
\refs{\hosono,\kachru}. In five dimensions, this model
contains $n_V=h_{(1,1)}-1=2$ vector multiplets (not counting the
graviphoton) and $n_H=h_{(2,1)}+1=244$
hypermultiplets.\foot{In this paper, we don't consider the hypermultiplet 
sector of $M$-theory where the low-energy effective action in $D=5$ 
does receive 
membrane and fivebrane instanton corrections \becker.}

Evidence for this identification from one-loop anomaly
tests will be shown below in section 4. For now, we simply note 
that, following \antft, it is straightforward 
to match the perturbative and non-perturbative BPS states arising 
from the ten-dimensional compactification with the states displayed 
in the previous section and arising
from the eleven-dimensional compactification.  

This can be seen as follows: from the ten-dimensional point of view,
the heterotic string compactified on $K3\times S^1$ has the perturbative 
fundamental string state with charge
\eqn\fundst{m_0=\int_{K3\times S^1\times S^2} H_7,}
where $H_7=e^{-\phi} *H_3$, $H_3$ is the field strength of the two-form
antisymmetric tensor field and $\phi$ is the ten-dimensional dilaton.
This state has mass per unit length $M_0=m_0g_5^2$. 
Here the string is reduced on, not
wrapped around, the $S^1$. The corresponding classical 
solution is given by the fundamental string of \dabghr.
This mass formula, 
which can be seen from central charge/supergravity considerations 
\antft, can also be
obtained by computing the ADM mass of the fundamental string 
solution. This state is associated with the $b_{\mu\nu}$ field and is
dual to a vector in $D=5$. 

The string theory also possesses
a perturbative electrically charged
point-like $H$-monopole (dual to the magnetically charged $H$-monopole state 
of \hmono) state with charge
\eqn\phmono{e_1=\int_{K3\times S^3} H_7,}
and with mass $M_1=e_1Rg_5$, where $R$ is the radius of the $S^1$ and $g_5$
is the five-dimensional string coupling constant. 
In this case, the string is wrapped around the $S^1$.
Again one obtains the
same mass from either the central charge or the ADM mass of the solitonic
solution. This state is associated
with the $b_{\mu 6}$ field. The $T$-dual electrically charged point-like 
Kaluza-Klein state with charge $e_2$ and associated with the 
$g_{\mu 6}$ field has mass $M_2=e_2g_5/R$. In this case, the corresponding
electrically charged solution is given by the extremal Kaluza-Klein black
hole solution of heterotic string theory \dufkmr.

The fundamental
string state can be identified with one of the three states shown
in the previous section arising from
the $M$-theory fivebrane, while the $H$-monopole and Kaluza-Klein states
can be identified with two of the three states shown in the previous
section arising from the $M$-theory membrane.   
 
The dual case is similar: 
the heterotic fivebrane wrapped around $K3\times S^1$ has the nonperturbative 
(from the string point of view) point-like state with charge
\eqn\fundpt{e_0=\int_{S^3} H_3,}
and mass $M'_0=e_0/g_5^2$ \refs{\wit,\antft}. Here the classical solution is 
simply the heterotic fivebrane of \duflufb\ wrapped around $K3\times S^1$,
and which is dual to the fundamental heterotic string.

One also gets from the heterotic
fivebrane a nonperturbative magnetically charged
string-like $H$-monopole state with charge
\eqn\shmono{m_1=\int_{S^1\times S^2} H_3,}
and mass per unit length $M'_1=m_1R/g_5$, where here the fivebrane is
wrapped around the $K3$ but reduced on the $S^1$.
The solution in this case is the usual magnetically
charged $H$-monopole, which in $D=5$ is a string \hmono.
The $T$-dual magnetically charged string-like 
Kaluza-Klein state with charge $m_2$ has mass per unit length
$M'_2=m_2/g_5R$. 

The point-like state can be identified with one of the three states shown
in the previous section arising from
the $M$-theory membrane, while the string-like $H$-monopole and Kaluza-Klein states
can be identified with two of the three states shown in the previous
section arising from the $M$-theory fivebrane\foot{In particular, since the point-like state coming from the heterotic fivebrane does not
come from the $M$-theory fivebrane, it follows that the two fivebranes
are not identical.}. 

Note that each of the 
three pairs of electric/magnetic dual states obey Dirac quantization 
conditions. Note also that neither the membrane nor the fivebrane from
$M$-theory is in itself sufficient to reproduce the perturbative spectrum
of either the five-dimensional string or the dual five-dimensional point-like
object. This becomes clear when one realizes that from the $M$-theory side,
the membrane wrapped around a two-cycle yields only point-like states, while
the fivebrane wrapped around a four-cycle yields only string-like states. On
the other hand, from the heterotic compactification,
both the string and point-like theories in $D=5$
contain both string and point-like objects in their perturbative spectra.
In particular, it follows that the $D=5$ spectrum of Calabi-Yau string
solitons yields the fundamental string states on the heterotic side as well
as the non-perturbative heterotic string states obtained by wrapping the 
heterotic fivebrane on $K3$. 

In reducing further to four-dimensions, one obtains the standard (point-like)
electric/magnetic duality. This entails wrapping the string around another
$S^1$ and reducing the point-like theory on $S^1$.

This four-dimensional duality can also be seen to arise directly from 
type $IIA$ membrane/fourbrane duality. We first reduce the membrane
of $M$-theory on $S^1$ to get the type $IIA$ membrane theory and
then compactify to four-dimensions on a Calabi-Yau manifold by wrapping
the membrane around a two-cycle. To get the dual point, we wrap the 
fivebrane of $M$-theory around $S^1$ to get the type $IIA$ fourbrane theory 
and then compactify on a Calabi-Yau manifold by wrapping the fourbrane around 
a four-cycle.

The connections between the fundamental states of the various
theories are shown in Fig.1.

\goodbreak\midinsert
\centerline{\epsfxsize 4.5truein\epsfbox{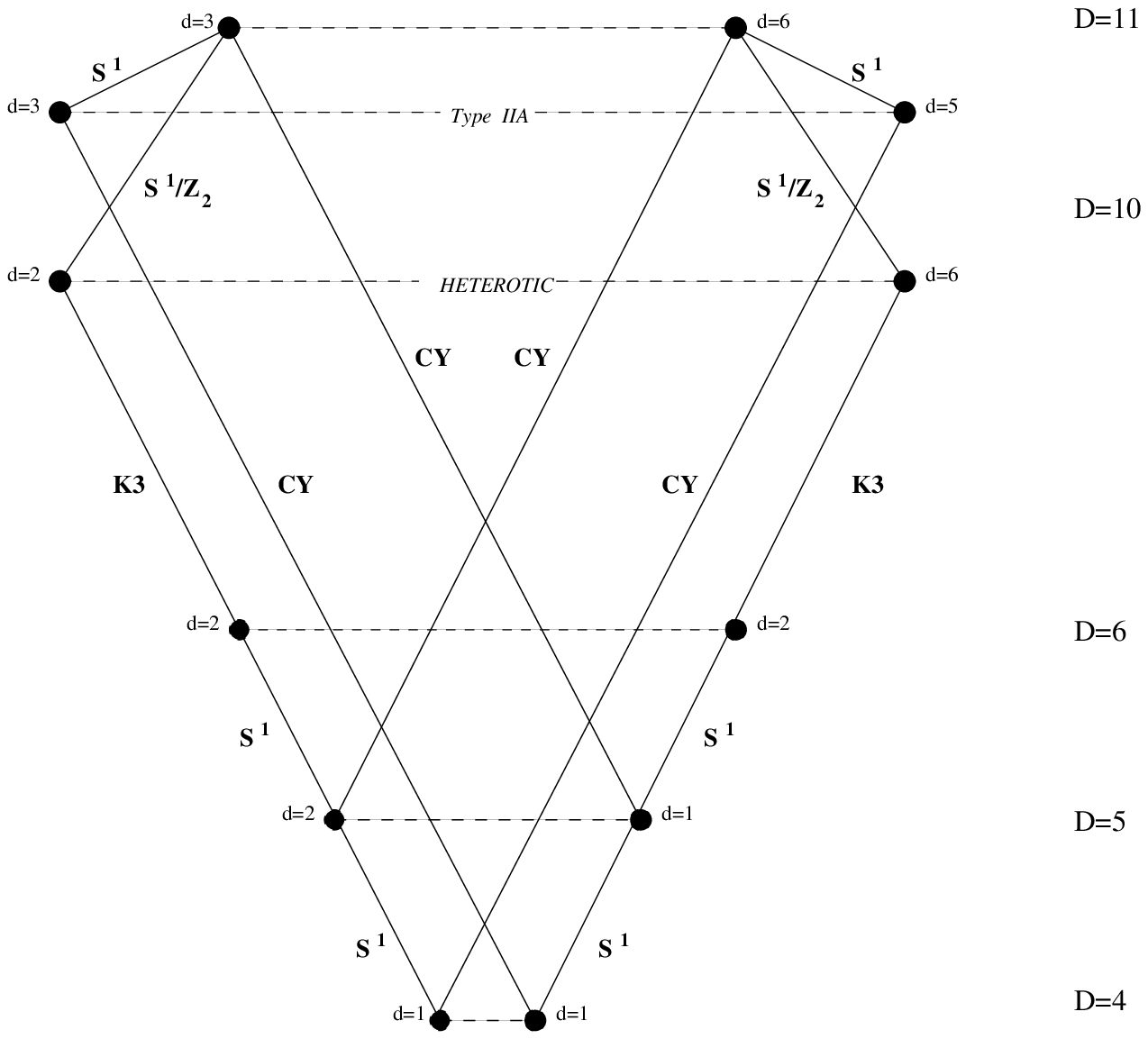}}
\smallskip
\leftskip 2pc \rightskip 2pc
\noindent{
\ninepoint\sl \baselineskip=8pt
{\bf Fig.1}
The ($N=2$) ``duality diamond''. The string and supergravity theories
connected by lines in the diagram possess fundamental states identified
under compactification. For example, the $D=5$ fundamental string is 
obtained either from the heterotic string theory reduced on $K3\times S^1$
or from the $M$-theory fivebrane wrapped around a Calabi-Yau four-cycle. Note,
however, that the five-dimensional heterotic string theory possesses
point-like perturbative states which come from the $M$-theory membrane.}
\smallskip\endinsert

\newsec{One-Loop Results}

The action of the eleven-dimensional supergravity limit of $M$-theory
is given by
\eqn\elevsg{I_{11}={1\over 2} \int_{M^{11}} d^{11}x \sqrt{-g}
\left[R - {1 \over 2} F_4 \wedge *F_4 - {1 \over 6} A_3 
\wedge F_4 \wedge F_4\right].}
This action should be augmented by a term predicted by membrane/fivebrane duality \dufflm
\eqn\newel{I_{11}^{Lorentz} = \int_{M^{11}} A_3\wedge
{1\over (2\pi)^4}\left[-{1\over 768}(\tr R^2)^2+{1\over 192}\tr R^4\right].}
The gravitational constant and the membrane and
fivebrane tensions are set to one.\foot{For a complete 
discussion of the tension quantization in $11$ dimensions see \dufflm.}
The reduction of \elevsg\ to five dimensions is well known 
(see, {\it e.g.}, \refs{\cadcdf,\antft}), and in addition to $h_{(1,1)}$
vectors and $h_{(2,1)}+1$ hypermultiplets has a topological coupling term \gst
\eqn\topcop{I_5^{top}=- {1 \over 12}C_{\Lambda\Sigma\Delta}\int_{M^5} 
A_1^\Lambda \wedge F_2^\Sigma \wedge F_2^\Delta.}
The $U(1)$ fields are normalised so that 
they couple to integer charges. On the other hand, the reduction of
\newel\ yields an interaction of the form
\eqn\new{I_5^{Lorentz} \sim \int_{M^5} \alpha_{\Lambda} A^\Lambda_1 \wedge 
\tr R^2,} where
\eqn\alp{\alpha_\Lambda={1\over 16(2\pi)^2}\int_{M^6} \omega_\Lambda \wedge \tr R_0^2,}
where $\Lambda=1,...,h_{(1,1)}$ and $\omega_\Lambda \in h_{(1,1)}$. The $\alpha_\Lambda$ 
define the topological couplings, which for $X_{24}(1,1,2,8,12)$ are
$(24, 48, 92)$ (see, {\it e.g.}, \hly).

Thus we obtain a set of $h_{(1,1)}$ vector equations of motion
\eqn\bian{d(G_{\Lambda\Sigma}H^\Sigma_3)=-{1\over 4}
\left[C_{\Lambda\Sigma\Delta} dA^\Sigma_1  dA^\Delta_1 +{1\over 24} 
\alpha_\Lambda \tr R^2\right],}
where $\Lambda,\Sigma,\Delta=1,\ldots h_{(1,1)}$ and 
$H_3^\Sigma=*F_2^\Sigma$. We follow the 
conventions 
of \cadcdf\ in defining the metric $G_{\Lambda \Sigma}$ and intersection
constants $C_{\Lambda \Sigma \Delta}$.
As explained in \antft, when the Calabi-Yau
manifold is a $K3$ fibration, one of these vectors can be dualized to give
a two-form field that can be identified with the $b_{\mu\nu}$ field
of the heterotic string on $K3\times S^1$. In the previous section, this
claim was supported at the level of BPS states. \bian\ can be obtained from 
the fivebrane (tree-level) Bianchi identity, involving gravitational 
Chern-Simons 
corrections arising from a sigma-model anomaly on the fivebrane worldvolume,
$dG_7=-{1\over2}F_4{}^2 + (2\pi)^4 {\tilde X}_8$ by decomposing
the fields in the basis of cohomology on the Calabi-Yau manifold. From 
the heterotic point of view, we see that the fivebrane Bianchi identity 
yields the string
Bianchi identity, involving the $b_{\mu\nu}$ field (tree-level), and 
$h_{(1,1)}-1$ vector equations of motion (one-loop effect).

As a further test, let us compare the holomorphic functions $F_1$ for the 
heterotic string and for $M$-theory for the specific three-moduli CY 
manifold. In the $M$-theory case, the absence in the 
$D=5$ spectrum of a scalar field corresponding to the
two-form antisymmetric tensor with both internal indices implies that there
are no non-perturbative corrections to the low-energy action describing the
vector multiplet and in particular its gravitational coupling:
\eqn\ffunc{F^M_1=24A_1^1+48A_1^2+92A_1^3.}
This can be viewed as the decompactification limit of the
four-dimensional type $II$ topological function \bers\
\eqn\ffour{F^{II}_1 ={12\pi i\over 12} c_2 \cdot \vec J+
non-perturbative \ \ corrections,}
where $c_2 \cdot \vec J = \alpha_{\Lambda} t^{\Lambda}=24t_1 + 48t_2 + 92t_3$ and where the non-perturbative corrections are absent in $D=5$.

On the heterotic side, the expression for our three moduli case is given
by \refs{\cardtwo,\antgnt}
\eqn\ffunch{F^{het}_1 = 24S_{inv}+ {2\over 4 \pi^2} \log(j(T)-j(U))
-{b_{grav}\over 8 \pi^2} \log\eta^{-2}(T)\eta^{-2}(U),}
where $b_{grav}=2(n_H-(n_V+1))+46=528$ for
$n_H=244$ and $n_V+1=3$.
$j(T)$ is the modular $j$-function, $\eta$ is the Dedekind function and 
$S_{inv}=(1/4\pi)S$. In the large $T$ limit this reduces to
\eqn\fhet{F^{het}={1\over 4\pi}\left(24S+48T+44U\right).}

Employing the connections between heterotic and $M$ moduli in the large moduli
limit $t_1=-iS; t_2=-i(T-U); t_3=-iU$ \cardoso\ ($T>U$ is assumed), 
one finds agreement in the large $T$ limit between the tree-level
$M$-theory result \ffunc\ and the one-loop heterotic expression \fhet\ 
for the three moduli case.
Agreement between the heterotic and type $IIA$ holomorphic functions
for the particular Calabi-Yau threefold $X_{12}(1,1,2,2,6)$ was found in
\kap. This model does not, however, arise from six-dimensional
heterotic/heterotic duality.

\newsec{Anomaly Cancellation from Strings}

It was pointed out in \refs{\town,\witfb}\
that, in the presence of a fivebrane, a
term representing the coupling of an anti-self dual three-form
field strength $T_3$ on the fivebrane worldvolume is necessary to cancel 
the anomaly from the interaction $\int_{M^{11}} A_3\wedge F_4 \wedge F_4$. 
This can be seen as follows. In the presence of a fivebrane with charge $m$,
\eqn\fbsource{dF_4=m\delta_V,}
where $\delta_V$ is supported on the fivebrane worldvolume $V$
({\it i.e.} it integrates to $1$ on the space transverse to the fivebrane).
So, under 
$\delta A_3=d\Lambda_2$,
\eqn\varbulk{\eqalign{{1\over 12}
\delta\left(\int_{M^{11}} A_3\wedge F_4 \wedge F_4\right)
&={1\over 4}\int_{M^{11}} d\Lambda_2\wedge F_4 \wedge F_4\cr
&=-{m\over 2} \int_V \Lambda_2\wedge F_4.\cr}}
This anomaly needs to be cancelled by a term
\eqn\sdtensor{{m\over 2}\int_V T_3\wedge A_3,}
where $T_3$ is the anti-self dual field three-form 
strength on the fivebrane worldvolume
and $dT_3=F_4$. 

An analogous situation arises for the five-dimensional theory in the presence
of string sources with charge
\eqn\stsource{dF_2^\Sigma=m^\Sigma\delta_W,}
where $\delta_W$ is supported on the string worldsheet.
The topological term 
$I_5^{top}=(-1/12)C_{\Lambda\Sigma\Delta}\int_{M^5} A_1^\Lambda
\wedge F_2^\Sigma \wedge F_2^\Delta$
is anomalous under $\delta A_1^\Lambda=d\lambda^\Lambda$:
\eqn\vartop{\eqalign{\delta I_5^{top}&=-{1\over 4}
C_{\Lambda\Sigma\Delta}\int_{M^5} d\lambda^\Lambda\wedge F_2^\Sigma 
\wedge F_2^\Delta\cr
&={m\over 2}^\Lambda C_{\Lambda\Sigma\Delta}\int_W \lambda^\Sigma\wedge 
F_2^\Delta.\cr}}
Another way of seeing this is to note that due to \stsource,
\bian\ is inconsistent: taking an external derivative
makes the left hand side vanish, while the right hand side is
nonzero.
The remedy is to add to the action a term
\eqn\sdvector{{1 \over 2} m^\Lambda C_{\Lambda\Sigma\Delta}\int_W 
T_1^\Sigma \wedge A_1^\Delta,}
where $T_1^\Sigma$ is a self-dual one-form field 
strength on the string worldsheet
and $dT_1^\Delta=F_2^\Delta$. This term cancels the $U(1)^{h_{(1,1)}}$
gauge anomalies of the bulk action in the presence of strings and arises 
as a part of a string worldsheet action analogous to
$D$-brane action\foot{In eleven dimensions, fivebranes can be 
interpreted as $D$-branes of open membranes \town. 
After compactification,
this picture reduces to point-like intersections of strings in five
dimensions.}  
presented in \town \eqn\sws{I_2 = {1\over 4}
d_{\Sigma\Delta}\int_W 
\left(T^\Sigma -{\hat *}A^\Sigma\right) \wedge \left({\hat *}T^\Delta - 
A^\Delta\right),}
where $d_{\Sigma\Delta}=C_{\Lambda\Sigma\Delta} m^\Lambda$;  here 
$A^\Sigma$ denote the pullbacks to the worldsheet of spacetime vectors 
and $\hat *$ is the dualization on the worldsheet.

Similarly, as in $D=11$ \refs{\dufflm, \witfb}, the interaction term of 
the form \new\ $\int \alpha_{\Lambda} F^\Lambda_2 \wedge \Omega_3$ which is
covariant in the 
absence of strings, now develops an anomaly due to \stsource:
\eqn\anommm{ \delta I_5^{Lorentz} = \alpha_{\Lambda} m^{\Lambda} 
\int_W \epsilon R,}
where $\epsilon$ is the infinitesimal parameter of the 
diffeomorphism 
(as a reminder, $\tr R^2=d\Omega_3$ and $\delta \Omega_3=d(\epsilon R)$).
Again, the worldsheet anomaly and the one in the interaction in the bulk 
are expected to cancel.\foot{In presence of a string source \stsource\ 
coming from the compactification of a fivebrane source on a CY manifold, 
supersymmetry is also anomalous \witcy. We assume that the supersymmetry
anomaly may be cancelled by a corresponding anomalous variation of the
string worldsheet action. This non-invariance may still require additional
terms both in the bulk lagrangian and in the worldsheet action.}
 It is clear from this consideration that the 
resulting five-dimensional string is necessarily chiral on the 
worldsheet.\foot{Another way of seeing this is to note, following
Witten \witfb, that in 
$D=5$, introducing a string worldsheet given by $x_3=x_4=x_5=0$ breaks
half of the spacetime supersymmetries and the surviving supersymmetries
obey $\Gamma_3\Gamma_4\Gamma_5=1$ or, equivalently, in view of
$\Gamma_1\Gamma_2\Gamma_3\Gamma_4\Gamma_5=1$, $\Gamma_1\Gamma_2=1$.}
This fact is also supported by the possible identification with the heterotic 
string compactified on $K3 \times S^1$ (for the suitable Calabi-Yau's);
from section 4 it can be seen that obtaining \anommm\ from the heterotic 
side requires both a tree-level and a one-loop calculation. A detailed
calculatioin of the anomalous worldsheet action of the string 
excitations of $M$-theory will be given elsewhere. 

We see that the five-dimensional theory mimics its eleven-dimensional 
``ancestor''
in many ways, at the same time having the advantage of being coupled 
to only string and point-like objects. Thus more detailed study of these 
five-dimensional theories may help in understanding $M$-theory while 
allowing calculations to be carried out in the more familiar setting of 
string theory. Finally, while further reduction on a circle is fairly 
straightforward and yields $N=2$ supersymmetric theories in $D=4$, 
as displayed in Fig.1, one may hope to obtain dual $N=1$ chiral theories 
following \duffmw\ by considering two different limits of $M$-theory 
compactified on $CY \times S^1/Z2$.

\vskip1truecm

\noindent
{\bf Acknowledgements:}

We would like to thank Gabriel Cardoso, Luciano Girardello and Camillo 
Imbimbo for helpful discussions. RK and RM were supported by World 
Laboratory Fellowships. One of us (SF) would like to thank the 
Institute for Theoretical Physics at Santa Barbara for its
kind hospitality and where part of this work was completed.

\vfil\eject
\listrefs
\bye